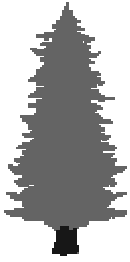

**2005 ALCPG & ILC Workshops – Snowmass, U.S.A.**

# Progress Towards a Long Shaping-Time Readout for Silicon Strips

Jurgen Kroseberg, Alex Grillo, Forest Martinez-McKinney, Gavin Nesom, Bruce A. Schumm, Ned Spencer, Max Wilder
*Santa Cruz Institute for Particle Physics, Santa Cruz, CA 95064, USA*

We report on progress in the development of the LSTFE2 prototype front-end ASIC geared towards the readout of silicon microstrip sensors in a Linear Collider detector. We also discuss progress in the design of the back-end digital architecture for this readout scheme.

The International Linear Collider (ILC) community currently has three detector designs under consideration: two based primarily on gaseous central tracking (LDC, GLD) and one that would rely entirely on silicon microstrips (SiD). Even in the case of the two TPC-based designs, though, the designs incorporate substantial intermediate and forward tracking systems based on silicon microstrip sensors. Thus, the development of a microstrip readout tailored to the conditions of the ILC is a high priority for the ILC detector R&D program.

The success of the ILC physics program is predicated on the precise measurement of the properties of new states such as the Higgs and supersymmetric partners of fundamental fermions and bosons. The front-end electronics for ILC detector microstrips must be designed with this in mind. This motivates several aspects of the readout development. In order to reduce the material burden of the tracker, and thus preserve the precision of the track-parameter resolution for low- and intermediate-momentum tracks, the readout should be able to turn on and off in less than a millisecond, in order to fully exploit the $5\times10^{-3}$ ILC duty cycle and eliminate the need for active cooling. The readout should also introduce as little noise as possible, allowing for the instrumentation of long single ladders for large-area applications, as well as for the most precise space-point resolution for conventional ladders. Finally, the readout should allow for timing on the order of the 337 nsec bunch separation of the ILC beam pulses.

We report here on the design and preliminary tests of the eight-channel LSTFE2 prototype microstrip readout ASIC, which was fabricated in the TSMC 0.25μm mixed-signal RF process, and was received at SCIPP on August 9, 2005. The LSTFE2 was designed with a 3μs shaping time – long enough to shape away much of the 1/f noise, but with the capability of providing a time resolution of better than 500 ns. Each LSTFE2 channel features two independent comparators: a high-threshold comparator to maintain a noise occupancy of $10^{-3}$, and a low-threshold comparator to allow the use of smaller pulse-height information in strips neighboring a high-threshold transition, to allow for improved centroid reconstruction. The LSTFE2 employs a high gain of 140 mV/fC, emphasizing the precision of pulse-height determination in the minimum-ionizing pulse height regime, as well as minimizing the effect of process variation in the comparators. Finally, the LSTFE2 ASIC includes control lines for the rapid cycling of the chip's power.

Preliminary tests of the LSTFE2 chip show that its major component blocks are working well. Figure 1 shows a confirmation of the 140 mV/fC gain for the combined preamplifier and shaper stages. The chip requires approximately 25 msec to reach operating condition after restoration of power, somewhat worse than the 1 msec goal, due to the fact that leakage currents erode latched bias levels after the power is turned off. It is suspected that channel protection diodes are the source of the leakage current; studies are underway to confirm this. A characterization of chip noise levels is also underway.

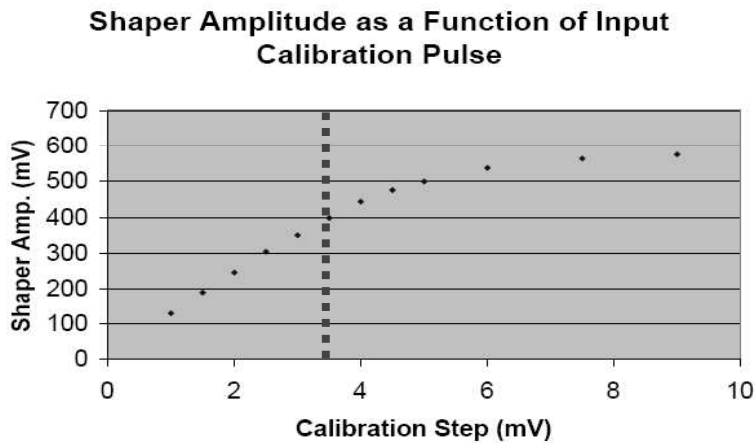

Figure 1: LSTFE2 combined preamplifier and shaper response to a calibration step, with a scale of approximately 1.1 fC per mV. The dashed blue line represents a charge injection equivalent to that of a 1 mip signal in a 300 μm thick silicon microstrip detector.

In collaboration with SCIPP radiology efforts, we have designed a prototype back-end architecture for the LSTFE chip. This architecture, to be implemented initially on an FPGA, employs logic that enables the low-threshold readout for channels in the vicinity of a channel that crosses the high threshold. For channels in this enabled region that cross the threshold, leading- and trailing-edge time stamps are written into a FIFO that serves the entire chip (8 channels for now; to be expanded to 128 channels in the next submission). This FIFO is then read out at high speed at the end of the 1ms pulse train.

Figure 2 shows a projection of the resulting data rate, assuming 0.1% noise occupancy and expected machine backgrounds in the innermost SiD tracking layer, per 1ms beam spill, for a 128 channel chip. At a repetition rate of 5 Hz, this corresponds to a data rate of approximately 35 kHz per 128 channel chip. Depending on whether long or short strips are implemented, proposed detector designs will have between 1-10 million channels, resulting in a relatively minimal overall data rate of order 0.5-5 GHz for the entire microstrip system.

Testing continues to determine chip characteristics such as noise levels as a function of load capacitance, temporal resolution, and performance when reading out a physical ladder in excess of 1 m (under construction at SCIPP). We are also employing a focused ion beam (FIBing) service to provide post-production modification of the fast turn-on circuitry, to help us identify the source of the leakage current that compromises turn-on performance.

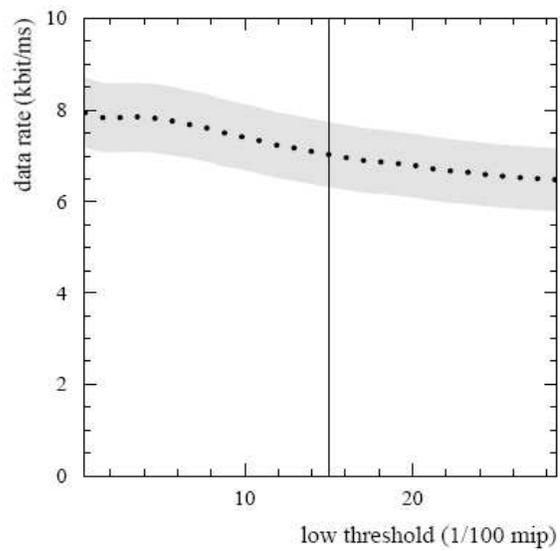

Figure 2: LSTFE data rate (in kbit per pulse train) per 128 channel chip, as a function of the lower threshold.

We plan to submit another prototype ASIC design (LSTFE3) in Spring 2005. In this design, we hope make use of what we learn about the source of leakage currents to protect against them, thereby approaching the desired turn-on specification. We also plan to further optimize the noise performance of the circuitry, and expand the chip to 128 parallel channels. We hope to explore the properties of the refined LSTFE3 front-end ASIC in a test-beam run towards the end of 2006.